\documentclass[12pt,letterpaper]{JHEP3}
\usepackage{graphics}
\usepackage{epsfig}

\title{The black hole dynamical horizon and generalized second law of thermodynamics}
\author{Song He\\
School of Physics, Peking University, Beijing, 100871, China\\
\email{she@pku.edu.cn}}
\author{Hongbao Zhang\\
Perimeter
Institute for Theoretical Physics, Waterloo, Ontario, N2L 2Y5,
Canada
\\Department of Astronomy, Beijing Normal University, Beijing,
100875, China\\
\email{hzhang@perimeterinstitute.ca}}

\abstract{The generalized second law of thermodynamics for a system
containing a black hole dynamical horizon is proposed in a covariant
way. Its validity is also tested in case of adiabatically collapsing
thick light shells.}


\begin{document}
\section{Introduction}
The generalized second law of thermodynamics was initially put forth
for a system including black holes by
Bekenstein\cite{Bekenstein1,Bekenstein2,Bekenstein3}. It states that
the sum of one quarter of the area of the black hole's event horizon
plus the entropy of ordinary matter outside never decreases with
time in all processes. It is noteworthy that for the formation or
absorption of black holes the generalized second law of
thermodynamics can also be equivalently formulated as a covariant
entropy bound. Namely, the entropy flux $S$ through the event
horizon between its two-dimensional space-like surfaces of area
$A_e$ and $A_e'$ must satisfy
\begin{equation}\label{eh}
S\leq\frac{A_e'-A_e}{4},
\end{equation}
where $A_e'\geq A_e$ is assumed.

However, due to the global and teleological property of event
horizon, the notion of dynamical horizon was developed and its
properties were investigated, where, in particular, the first and
second laws of black hole mechanics was generalized to the dynamical
horizon\cite{AK0,AK1,AK2,AG}. Thus it is tempting to conjecture that
the dynamical horizon may also have the thermal character as the
event horizon does, and the generalized second law of thermodynamics
may also be applied to the dynamical horizon. This is what we shall
address in the present paper. In next section, we shall propose a
covariant entropy bound formulation of the generalized second law of
thermodynamics associated with the black hole dynamical horizon.
Then its validity is demonstrated in a model of a growing black hole
by spherically symmetrical collapse of thick light shells. Some
discussions are presented in the end.

The signature of metric takes $(-,+,+,+)$. Notation and conventions
follow Ref.\cite{Wald}.
\section{The generalized second law of thermodynamics associated with the black hole dynamical horizon\label{definition}}
We would first like to introduce the basic definition of the black
hole dynamical horizon. For more subtle details, please refer to
Ref. \cite{AK1} and references therein.

\emph{Definition}: A smooth, three-dimensional, space-like
sub-manifold in a space-time ($M, g_{ab}$) is said to be a black
hole dynamical horizon if it can be foliated by a family of closed
two-dimensional surfaces such that, on each leaf, the expansion
$\theta_l$ of one future-directed null normal $l^a$ vanishes and the
expansion $\theta_n$ of the other future-directed null normal $n^a$
is strictly negative. If we choose the normalization of $l^a$ and
$n^a$ such that $l^an_a=-2$, then the expansion of the null
geodesics normal can be given by
$\theta_l=h^{ab}\nabla_al_b$($\theta_n=h^{ab}\nabla_an_b$) with the
induced metric $h_{ab}=g_{ab}+\frac{1}{2}(l_an_b+n_al_b)$ on each
leaf.

\begin{figure}[htb!]
\hspace{+.23\textwidth} \vbox{\epsfxsize=0.5\textwidth
  \epsfbox{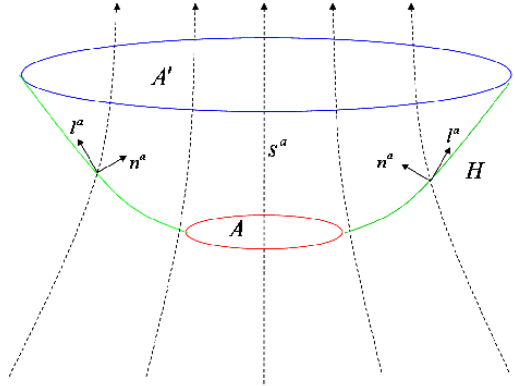}}
\caption {\small\sl A black hole dynamical horizon $H$ between its
apparent horizon of area $A$ and $A'$ with entropy current $s^a$
flowing through it.} \label{dh}
\end{figure}

Thus, roughly speaking, a black hole dynamical horizon is a
space-like hyper-surface which is foliated by closed apparent
horizons, where $l^a$ and $n^a$ represent future-directed outgoing
and ingoing null normals, respectively. See Fig.\ref{dh}. Note that,
in contrast to the notion of the event horizon, the dynamical
horizon can be identified quasi-locally without knowledge of the
full space-time history. In addition, intuitively, it is clear that
no signal can propagate out of the dynamical horizon due to the fact
that the dynamical horizon is space-like. All of these make the
dynamical horizon become a competent candidate for the boundary of
the black hole.

Now associated with the alternative boundary of the black hole, the
generalized second law of thermodynamics can be naturally formulated
in a covariant way as follows: \emph{The entropy flux $S$ through
the black hole dynamical horizon between its apparent horizons of
area $A$ and $A'$ must satisfy $S\leq\frac{A'-A}{4}$ if the dominant
energy condition holds for matter, where $A'>A$ is assumed.}

In the subsequent section, its validity will be tested by
adiabatically collapsing thick light shells.
\section{The generalized second law of thermodynamics tested by adiabatically collapsing thick light shells}
Start with a model of formation of a black hole by spherically
symmetrical collapse of thick light shells in Eddington-Finkelstein
coordinate\cite{ABCL}
\begin{equation}
ds^2=-dt^2+dr^2+r^2d\Omega^2+u(dt+dr)^2,
\end{equation}
where\begin{eqnarray}\label{universal}
 u = \left\{ \begin{array}{ll}
0, &  \textrm{Minkowski region},  \\
\frac{2m}{r}F(\frac{r+t}{\Delta}), & \textrm{within light shells},\\
\frac{2m}{r}, & \textrm{Schwarzschild region}.
\end{array} \right.
\end{eqnarray}
Here both $m$ and $\Delta$ are constant parameters. In addition, $F$
is a monotonically increasing function with respect to its argument,
with $F(0)=0$, and $F(1)=1$. Thus $F$ essentially serves as a
density profile function. After a straightforward but tedious
calculation, one finds that this metric is a solution of Einstein
equation with the non-vanishing energy momentum tensor within light
shells being given by
\begin{equation}\label{em}
T_{ab}=\frac{mF'}{4\pi\Delta r^2}k_ak_b,
\end{equation}
where $F'$ denotes the derivative of $F$ with respect to its
argument, and the null vector field $k_a=(dt)_a+(dr)_a$. Clearly,
the energy momentum tensor satisfies the dominant energy condition
due to $F'>0$.

\begin{figure}[htb!]
\hspace{+.33\textwidth} \vbox{\epsfxsize=0.7\textwidth
  \epsfbox{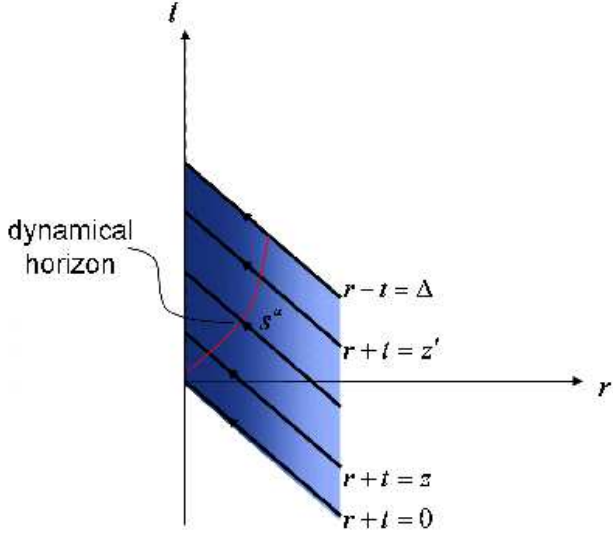}}
\caption {\small\sl a black hole is being formed by collapse of
thick light shells between $r+t=0$ and $r+t=\Delta$ in
Eddington-Finkelstein coordinate, falling through the black hole
dynamical horizon $r=2mF$, which also serves as the infinite
redshift surface.} \label{gc}
\end{figure}

Specifically speaking, this model describes a family of concentric
light shells with a flat Minkowski interior, ending with a final
black hole of Schwarzschild radius $R=2m$. The innermost light shell
reaches the center at the time $t=0$, and after the total duration
of collapse $\Delta$ the outmost light shell finally arrives at the
singularity. See Fig.\ref{gc}.

Next, to locate the black hole dynamical horizon in this model, let
us first compute the initial expansion of the future-directed null
normal to an arbitrary sphere characterized by some value of
$(t,r)$. The outgoing and ingoing null normals to these spheres can
be chosen to be, respectively,
\begin{equation}
l^a=(1+u)(\frac{\partial}{\partial
t})^a+(1-u)(\frac{\partial}{\partial r})^a,
n^a=k^a=(\frac{\partial}{\partial t})^a-(\frac{\partial}{\partial
r})^a,
\end{equation}
then the corresponding expansions can be obtained as
\begin{equation}
\theta_l=\frac{2(1-u)}{r},\theta_n=-\frac{2}{r}.
\end{equation}
Obviously, it follows from the definition presented in
Sec.\ref{definition} that the hyper-surface $u=1$ is a black hole
dynamical horizon if and only if its normal vector field is
time-like, i.e.,
\begin{equation}
g^{ab}\nabla_au\nabla_bu|_{u=1}=2\frac{\partial u}{\partial
t}(\frac{\partial u}{\partial r}-\frac{\partial u}{\partial
t})|_{u=1}<0.
\end{equation}
Thus, according to Eq.(\ref{universal}), it is easy to find that the
black hole dynamical horizon here is only the hyper-surface $u=1$
within light shells, i.e., between the hyper-surfaces $r+t=0$ and
$r+t=\Delta$, as shown in Fig.\ref{gc}.

To proceed, we further assume that the collapse of light shells is
adiabatical. Therefor the conserved entropy current of light shells
can be written as
\begin{equation}
s^a=\frac{s'(\frac{r+t}{\Delta})}{4\pi\Delta r^2}k^a,
\end{equation}
where the derivative of a function $s$ with respect to its argument
$s'>0$.

We shall now check whether the generalized second law of
thermodynamics is satisfied for the black hole dynamical horizon. As
demonstrated in Fig.\ref{gc}, let $z'>z$, then the area difference
of apparent horizons lying in $r+t=z$ and $r+t=z'$ reads
\begin{equation}\label{area}
\delta A=16\pi m^2[F^2(\frac{z'}{\Delta})-F^2(\frac{z}{\Delta})].
\end{equation}
On the other hand, by the conservation of the entropy current and
Gauss theorem, the entropy flux $S$ through the black hole dynamical
horizon between the above apparent horizon is equal to that through
the space confined within $z-t_p<r<z'-t_p$ at a time $t_p$ in the
distant past. Note that in the distant past, the light shells
resided in asymptotically flat region. Thus by Eq.(\ref{em}), the
effective mass of light shells between $r+t=z$ and $r+t=z'$ can be
obtained as
\begin{equation}
M_{eff}=m[F(\frac{z'}{\Delta})-F(\frac{z}{\Delta})],
\end{equation}
which equals the mass of the final black hole formed by collapse of
these light shells. So employing Eq.(\ref{eh}), we have an upper
bound on the entropy flux $S$
\begin{equation}\label{entropy}
S\leq 4\pi m^2[F(\frac{z'}{\Delta})-F(\frac{z}{\Delta})]^2.
\end{equation}
Combining Eq.(\ref{area}) with Eq.(\ref{entropy}), we have
\begin{equation}
S\leq\frac{\delta A}{4},
\end{equation}
which confirms the generalized second law of thermodynamics for the
black hole dynamical horizon.
\section{Discussions}
we have proposed a new generalized second law of thermodynamics
based on the notion of a black hole dynamical horizon. Its validity
has also been demonstrated in a physically reasonable model of black
hole formation by adiabatical collapse of thick light shells.

As mentioned in the beginning, along with the first and second laws
of black hole mechanics for the dynamical horizon , our result
further implies the black hole dynamical horizon may also have an
interpretation of thermodynamics, especially one quarter of area of
the black hole dynamical horizon may be identified with its entropy.
It is therefore interesting to analyze if a derivation of the black
hole entropy is available for the dynamical horizon based on the
counting of micro-states in quantum gravity such as causal set
theory, loop quantum gravity and string
theory\cite{DS,RZ,Ashtekar,Donnelly,SV}.

Even if it turns out that the black hole dynamical horizon has no
interpretation of thermodynamics in an underlying quantum theory of
gravity, our proposal can still be viewed as a covariant entropy
bound conjecture on the dynamical horizon. It is noteworthy that its
validity has also been verified in the cosmological context no
matter whether the dynamical horizon is space-like or not\cite{HZ}.
Thus it is natural to expect that our proposal as a covariant
entropy bound holds for the time-like analog of the black hole
dynamical horizon.

\acknowledgments HZ would like to give thanks to Professors Robert
B. Mann and Rafael D. Sorkin for their stimulating questions on the
dynamical horizon for the black hole being formed by collapse of
shells during his talk at Perimeter Institute, which speeds up this
work. He is also indebted to Yidun Wan for his clarifying the
essential difference between the event horizon and dynamical
horizon. Work by SH was supported by NSFC(nos.10235040 and
10421003). HZ was supported in part by the Government of China
through CSC(no.2007102530). This research was supported by Perimeter
Institute for Theoretical Physics. Research at Perimeter Institute
is supported by the Government of Canada through IC and by the
Province of Ontario through MRI.


\begin{thebibliography}{99}
\bibitem{Bekenstein1}J. D. Bekenstein, Nuovo Cim. Lett. 4, 737(1972).
\bibitem{Bekenstein2}J. D. Bekenstein, Phys. Rev. D7, 2333(1973).
\bibitem{Bekenstein3}J. D. Bekenstein, Phys. Rev. D9, 3292(1974).
\bibitem{AK0}A. Ashtekar and B. Krishnan, Phys. Rev. Lett. 89,
261101(2002)[arXiv:gr-qc/0207080].
\bibitem{AK1}A. Ashtekar and B. Krishnan, Phys. Rev. D68,
104030(2003)[arXiv:gr-qc/0308033].
\bibitem{AK2}A. Ashtekar and B. Krishnan, Living Rev. Rel. 7,
10(2004)[arXiv:gr-qc/0407042].
\bibitem{AG}A. Ashtekar and G. J. Galloway, Adv. Theor. Math. Phys.
9, 1(2005)[arXiv:gr-qc/0503109].
\bibitem{Wald}R. M. Wald, General Relativity(The University of Chicago
Press, Chicago, 1984).
\bibitem{DS}D. Dou and R. D. Sorkin, Found. Phys. 33, 279(2003)[arXiv:gr-qc/0302009].
\bibitem{RZ}D. Rideout and S. Zohren, Class. Quant. Grav. 23,
6195(2006)[arXiv:gr-qc/0606065].
\bibitem{Ashtekar}A. Ashtekar, J. Baez, A. Corichi, and K. Krasnov,
Phys. Rev. Lett. 80, 904(1998)[arXiv:gr-qc/9710007].
\bibitem{Donnelly}W. Donnelly, in private communication on entanglement entropy in loop quantum
gravity.
\bibitem{SV}A. Strominger and C. Vafa, Phys. Lett. B379.
99(1996)[arXiv:hep-th/9601029].
\bibitem{ABCL}R. J. Adler, J. D. Bjorken, P. Chen, and J. S. Liu, Am. J. Phys. 73, 1148(2005)[arXiv:gr-qc/0502040].
\bibitem{HZ}S. He and H. Zhang, JHEP 10,
077(2007)[arXiv:0708.3670].
%\bibitem{Bousso0}R. Bousso
%{\it{et al.}},Phys. Rev. D68
%064001(2003)[arXiv:hep-th/0305149].
%\bibitem{Bousso1}R. Bousso, Rev. Mod. Phys. 74, 825(2002)
%[arXiv:hep-th/0203101]
%\bibitem{Bousso2}R. Bousso, JHEP 9907,
%004(1999)[arXiv:hep-th/9905177].
%\bibitem{Bousso3}R. Bousso, JHEP 9906, 028(1999)[arXiv:hep-th/9906022].
%\bibitem{Bousso4}R. Bousso, Class. Quant. Grav. 17,
%997(2000)[arXiv:hep-th/9911002].
%\bibitem{Flanagan}E. E. Flanagan {\it{et al.}}, Phys. Rev. D62,
%084035(2000)[arXiv:hep-th/9908070].
\end{thebibliography}
\end{document}